\documentclass[12pt,a4paper]{article}

\usepackage[utf8]{inputenc}
\usepackage[T1]{fontenc}

\usepackage{amsmath,amssymb,amsfonts,bm}
\usepackage{mathtools}
\usepackage{physics}
\usepackage{graphicx}
\usepackage{float}
\usepackage{booktabs}
\usepackage{geometry}
\usepackage{caption}
\usepackage{subcaption}
\usepackage{siunitx}
\usepackage{multirow}
\usepackage{authblk}
\usepackage{hyperref}
\usepackage{microtype}
\usepackage{cleveref} 

\makeatletter
\def\@seccntformat#1{\hspace*{0pt}\csname the#1\endcsname.\quad}
\def\section{\@startsection{section}{1}{0pt}
{3.5ex plus 1ex minus .2ex}
{2.3ex plus .2ex}
{\centering\normalfont\large\bfseries}}
\makeatother

\title{\textbf{\boldmath HWF Analysis of TRN Black Holes}}

\author[]{Ashfaque Ahmad\thanks{gq0084@myamu.ac.in}}
\author[]{Sudhir Kumar Gupta\thanks{sudhir.ph@amu.ac.in}}
\author[]{Abbas Ali\thanks{aali.ph@amu.ac.in}}

\affil[]{Department of Physics, Aligarh Muslim University, Aligarh, UP -- 202002, India}

\date{} 

\hypersetup{pdftitle={HWF Analysis of TRN Black Holes},
pdfauthor={Ashfaque Ahmad, Sudhir K. Gupta, Abbas Ali},
unicode=true,
colorlinks=true,
linkcolor=blue,
citecolor=blue,
urlcolor=blue
}

\begin{document}

\maketitle

\begin{abstract}
	We extend the horizon wave-function (HWF) formalism to a spherically symmetric, charged, spinless quantum source in $\mathcal{D} + 1$ dimensional spacetime geometry and construct the HWFs associated with inner- as well as outer-horizons of the Tangherlini--Reissner--Nordstr\"om (TRN) black hole. The corresponding black hole formation probabilities, $P_{\rm BH}^{\pm}$, are also calculated and their sensitivities due to various source parameters including charge, mass, and $\mathcal{D}$ are analysed as well. Our analysis reveals that the ratio of inner- to outer-horizon probabilities, $P_{\rm BH}^{-}/P_{\rm BH}^{+}$, changes drastically from being unity as the source departs away from being in the extremal limit defined by $\alpha$ ($\to 1$ for extremal case). Furthermore, black hole formation probabilities are found to be highly suppressed for higher values of $\mathcal{D}$, particularly for the source mass, $m \lesssim \Lambda_{\mathcal{D}}$, e.g., for $\alpha = 0.5$, the ratio, $P_{\rm BH}^{+}(\mathcal{D} = 5)/P_{\rm BH}^{+}(\mathcal{D}=10)$ turn out to be approximately $2.95\times10^3$, $117$, and $3.35$ respectively for $m/\Lambda_{\mathcal{D}} = 0.5, 1$ and $2$, with $\Lambda_{\mathcal{D}}$ being the reduced gravity scale associated with $\mathcal{D}$ spatial-dimensions.
\end{abstract}

\clearpage
\section{Introduction} 
\label{sec:introduction}
In classical general relativity, a sufficiently compact distribution of energy can form a trapped surface and, under appropriate conditions, a black hole horizon. For a non-rotating, uncharged object in $(3+1)$ spacetime dimensions, the Schwarzschild radius is given as $R_S = {2GM}$ in natural units~\cite{Bel:2007zz, Schwarzschild:1916uq}. The classical criterion for black hole formation is given by the hoop conjecture~\cite{Thorne:1972ji, Ida:2002hg} according to which formation of a horizon requires compression of mass within a region with circumference below $2\pi R_S$ along all directions. However it has been established that the hoop conjecture cease to work while the source is a quantum object whose spatial localisation width is equal to its Compton wavelength~\cite{Casadio:2013aua}. In this regime the uncertainty in the quantum fluctuations associated with the source and horizon radius require probabilistic description of black hole formation. Several approaches have been developed to investigate the quantum nature of black holes, including self-completion~\cite{Dvali:2010bf} and N-portrait scenarios~\cite{Dvali:2011aa}, generalised uncertainty principles and black hole remnants~\cite{Adler:1999bu, Scardigli:1999jh, Hossenfelder:2012jw, Alberghi:2013hca}, and the gravitational collapse in high-energy wave-packet collisions~\cite{Giddings:2004xy}. In this work, we adopt the horizon wave function (HWF) formalism~\cite{Casadio:2013tma,Casadio:2013uga}, which provides a framework for describing horizon fluctuations and estimating the probability of black hole formation for a quantum source. Higher-dimensional black holes~\cite{Tangherlini:1963bw,Myers:1986un,Emparan:2008eg,Tomizawa:2011mc} have been well studied in the context of extra spatial-dimensional models, wherein the Planck scale can be much smaller than the $(3+1)$-spacetime Planck scale ($M_{Pl}$), potentially bringing gravitational phenomena within the reach of experiments~\cite{Meschini:2006gm}. In particular, the Arkani-Hamed-Dimopoulos-Dvali (ADD)~\cite{Antoniadis:1998ig, Arkani-Hamed:1998jmv} and Randall-Sundrum (RS)~\cite{Randall:1999ee, Randall:1999vf} models suggest that black holes could be produced in high-energy collisions~\cite{Argyres:1998qn, Banks:1999gd, Dimopoulos:2001hw, Gingrich:2007fm, Giddings:2001ih, Kanti:2004nr, Cavaglia:2002si,Dimopoulos:2001qe, Koch:2005ks} or in cosmic-ray interactions~\cite{Feng:2001ib, Anchordoqui:2001cg}. Classical trapped-surface analyses~\cite{Eardley:2002re, Yoshino:2002tx, Yoshino:2005hi} and semiclassical arguments~\cite{Voloshin:2001vs} give different expectations for the formation rate, and collider searches have so far found no signal of microscopic black holes~\cite{CMS:2026cgd, ATLAS:2026hzb,ATLAS:2023vat, Ahmad:2026hcy}. These considerations motivate a closer examination of quantum effects in black hole formation near the $\Lambda_{\mathcal D}$, where classical description of the source may not be sufficient. The HWF formalism addresses this issue in the following manner: Consider a massive source confined within a quantum wave packet whose spatial-width is of the order of its own gravitational horizon. In such a case, the position of horizon can not be regarded as a fixed classical quantity. The HWF treats the horizon radius as a quantum operator, linked to the momentum of source via the mass-shell condition. Instead of assuming that a horizon forms classically, one  calculates the probability that the quantum state of the source satisfies the conditions for the horizon to exist. This makes it possible to construct a probability distribution for the horizon radius and, consequently, to determine the probability that the source is contained within its own corresponding horizon radius. The formalism provides a general framework for the quantum fluctuations of the horizon and the probabilistic nature of black hole formation~\cite{Casadio:2015qaq, Casadio:2013tma, Casadio:2016fev, Casadio:2017nfg, Casadio:2018rlc, Arsene:2016kvf, Feng:2025nai, Spallucci:2021ljr}. The effects of charge and $\mathcal{D}$ spatial-dimensions have so far been studied separately within the HWF framework. The $(3+1)$-dimensional charged analysis for the Reissner--Nordstr\"om geometry~\cite{Casadio:2015rwa}, shows the black hole probabilities associated with the inner- and outer-horizons of massive charged sources. It is found that quantum-effects are particularly significant for the inner Cauchy horizon. For certain ranges of mass and charge, the inner-horizon probability is very small while the outer-horizon probability is close to one. HWF have been also extended for the Schwarzschild metric in $\mathcal{D}$ spatial-dimensions~\cite{Casadio:2015jha}, wherein for a fixed mass of the source, the probability of black hole formation decreases as $\mathcal{D}$ is increased~\cite{Arsene:2016kvf}, which indicates that the dimensionality of space is quite important for quantum black hole formation. However the combined effect of charge and dimensions has not been explored within the HWF framework. In the present work, we extend the HWF to a charged source in $(\mathcal{D}+1)$ dimensions, construct the HWFs of the outer- and inner-horizons of the Tangherlini--Reissner--Nordstr\"om geometry, and calculate the associated formation probabilities. This allows us to study how the charge parameter and higher dimensions affect the probability of black hole formation.

The organisation of the article is as follows: in Section~\ref{sec:source} we present the quantum source, a spherically symmetric Gaussian wave function in $\mathcal{D}$ spatial-dimensions, and its momentum-space representation, which provides the energy distribution needed to construct the HWF. Section~\ref{sec:charged_HWF_D} introduces the Tangherlini--Reissner--Nordstr\"om geometry and obtains the HWFs $\psi_H^{\pm}(r_\pm)$ by mapping the momentum-space distribution onto the two horizon branches. We then construct the joint probability densities and compute the formation probabilities $P_{\rm BH}^{\pm}$, studying their dependence on $\mathcal{D}$, the source mass $m$, the source width $\ell$ and the charge parameter $\alpha$. Finally, Section~\ref{sec:summary} summarises the results.

\section{Quantum Source in TRN Geometry}
\label{sec:source}
Following previous HWF in $\mathcal{D}$ spatial-dimensions~\cite{Casadio:2014twa, Casadio:2013tza}, we consider a spherically symmetric Gaussian wave function to represent the quantum state of a charged, non-rotating source with width $\ell$ at rest in $\mathcal{D}$ spatial-dimensions. The source wave function is written as~\cite{Casadio:2015jha},
\begin{equation}
\psi_S(r) = A_\mathcal{D}\exp\!\left(-\frac{r^2}{2\ell^2}\right).
\label{eq:gaussian_ansatz}
\end{equation}
Here, the radial variable $r$ is interpreted as the same coordinate $r$ as in the metric~\eqref{eq:metric} and $A_\mathcal{D}$ is a dimension-dependent normalisation constant. 
The $\mathcal{D}$-dimensional volume element is $d^{\mathcal{D}} r = r^{\mathcal{D}-1}\,dr\,\mathcal{D}\Omega_{\mathcal{D}-1}$ with $\Omega_{\mathcal{D}-1} = {2\pi^{\mathcal{D}/2}}/{\Gamma(\mathcal{D}/2)}$, where $\Omega_{\mathcal{D}-1}$ is the surface area of the unit $(\mathcal{D}-1)$ sphere and using these we obtain normalisation constant as, $A_\mathcal{D} = {1}/ ({\pi^{\mathcal{D}/4}\ell^{\mathcal{D}/2}})$. Consequently, the normalised source wave function turns out to be,  
\begin{equation}
\psi_S(r) = \frac{1}
{\pi^{\mathcal{D}/4}\ell^{\mathcal{D}/2}}
\exp\!\left(-\frac{r^2}{2\ell^2}\right).
\label{eq:normalised_source}
\end{equation}
The probability of finding the source in the spherical shell between $r$ and $r+dr$ also carries the geometric factor $\Omega_{\mathcal{D}-1}r^{\mathcal{D}-1}$, giving the radial probability distribution, 
\begin{equation}
\mathcal P_S(r)
=
\Omega_{\mathcal{D}-1} r^{\mathcal{D}-1}
\left|\psi_S(r)\right|^2
=
\frac{\Omega_{\mathcal{D}-1}}
{\pi^{\mathcal{D}/2}\ell^\mathcal{D}}
r^{\mathcal{D}-1}
\exp\left(-\frac{r^2}{\ell^2}\right).
\label{eq:PS_radial_density}
\end{equation}
We now calculate the probability of finding the source within a sphere of radius $r_\pm$. In order to obtain it, we integrate the radial probability density from the origin to the respective horizon,
\begin{equation}
P_S(r<r_\pm)
\equiv
\int_0^{r_\pm}
\mathcal{P}_S(r) \, dr
=
\Omega_{\mathcal{D}-1}
\int_0^{r_\pm}
\left|\psi_S(r)\right|^2
r^{\mathcal{D}-1} \, dr .
\label{eq:PS_rpm_definition}
\end{equation}
$P_S(r< r _ +)$ is the probability that the source is within the outer-horizon, while $P_S(r<r_-)$ is the probability that the source is within the inner-horizon. From Eq.~\eqref{eq:PS_rpm_definition} we get,

\begin{equation}
P_S(r<r_\pm)
=\frac{\gamma\left(\frac{\mathcal{D}}{2},
\frac{r_\pm^2}{\ell^2}
\right)}{\Gamma(\mathcal{D}/2)}, 
\label{eq:PS_rpm_final}
\end{equation}
where $\gamma(a,x)$ and $\Gamma(a,x)$ are the lower and upper incomplete gamma functions. 

As the HWF description requires source in momentum-space rather than position-space, we	 therefore, write the Fourier transformed analog of $\psi_S(r)$ as, 
\begin{equation}
\widetilde\psi_S(p)=\frac{1}{(\sqrt\pi\,\Delta_p)^{\mathcal{D}/2}}\exp\!\left(-\frac{p^2}{2\Delta_p^2}\right)
\label{eq:momentum_wavefunction}.
\end{equation}
Here $\Delta_p$ is the momentum width. The Fourier transform of a Gaussian of width $\ell$ has $\Delta_p = \hbar/\ell$. The localisation width of the source is determined by its $\mathcal{D}$-dimensional Compton wavelength $\lambda^\mathcal{D}_{c}$, where $\lambda^\mathcal{D}_{c}$ =$\ell$ = ${\ell_{\mathcal{D}}\Lambda_{\mathcal{D}}}/{m}$~\cite{Casadio:2014twa}. In natural units $\ell_\mathcal{D} \Lambda_\mathcal{D} = 1$ in terms of reduced Planck scale and Planck length scales in $\mathcal{D}$-dimensional.

\section{Horizon Quantum Mechanics and the TRN Black Hole}
\label{sec:charged_HWF_D}
We now turn our focus on finding the black hole formation probabilities within the HWF framework for the  Tangherlini--Reissner--Nordstr\"om (TRN) black hole which is described by the following metric, ~\cite{Tangherlini:1963bw, Emparan:2008eg, Tomizawa:2011mc}
\begin{equation}
ds^2=-f_\mathcal{D}(r)\,dt^2+f_\mathcal{D}(r)^{-1}dr^2+r^2\,\mathcal{D}\Omega_{\mathcal{D}-1}^2,
\qquad
f_\mathcal{D}(r)=1-\frac{\Sigma_\mathcal{D}}{r^{\mathcal{D}-2}}+\frac{Q^2}{r^{2(\mathcal{D}-2)}},
\label{eq:metric}
\end{equation}
The locations of the horizons are determine by the positive roots of $f_\mathcal{D}(r)=0$. Introducing the variables $x=r^{\mathcal{D}-2}$, this conditions can be written as $x^2-\Sigma_\mathcal{D}x+Q^2=0$. For real horizons to exist, the discriminant of this equation must be non-negative, which gives $\Sigma_\mathcal{D}^2\ge4Q^2$. For convenient, we consider the dimensionless specific charge, $\alpha = {2 Q}/{\Sigma_\mathcal{D}}={(\mathcal{D}-2)Q}/{G_\mathcal{D} M}$ with $\Sigma_\mathcal{D}={2G_\mathcal{D}M}/{(\mathcal{D}-2)}$ and $\mathcal{D}$-dimensional Newton constant given as $G_\mathcal{D}={\ell_\mathcal{D}^{\mathcal{D}-2}}/{\Lambda_{\mathcal{D}}}$, which gives as $G_3=\ell_P/m_P$ for $\mathcal{D}=3$. Here $M$ is the ADM mass and $Q$ is the charge of the source. The coefficient in $\Sigma_\mathcal{D}$ follows the convention of earlier HWF work~\cite{Casadio:2015rwa, Casadio:2015sda}. The two positive roots of $f_\mathcal{D}(r)=0$ are defined as the outer- and inner-horizon radii, such as
\begin{equation}
	R_\pm^{\mathcal{D}-2}=\frac{\Sigma_\mathcal{D}}{2}\,\beta_\pm,\qquad
	\beta_\pm(\alpha)=1\pm\sqrt{1-\alpha^2},
	\label{eq:horizons}
\end{equation}
where the parameters $\beta_\pm$ satisfy the relations $\beta_++\beta_-=2$ and $\beta_+\beta_-=\alpha^2$. The requirement that $\beta_\pm(\alpha)$ be real imposes the condition $1-\alpha^2\geq 0$, which implies $|\alpha|\leq 1$. Therefore allowed range is $\alpha\in[0,1]$, ensuring that both horizon radii are real. For $0\leq\alpha<1$, the spacetime admits two distinct horizons, namely the outer-horizon $R_+$ and the inner (Cauchy)-horizon $R_-$, with $R_-<R_+$. In the extremal limit $\alpha=1$, the two horizons coincide, $R_+=R_-$. For $\alpha>1$, the horizon radii become complex, and the singularity is no longer shielded by an event horizon, corresponding to a naked singularity. This regime is excluded from the present analysis. The corresponding horizon radii in terms of $M$ are obtained by substituting the values of $\Sigma_\mathcal{D}$ and $G_\mathcal{D}$ into Eq.~\eqref{eq:horizons}, we get
\begin{equation}
R_{\pm}(M)=\ell_\mathcal{D}\left[\frac{\beta_{\pm}\,M}{(\mathcal{D}-2)\,\Lambda_{\mathcal{D}}}\right]^{\frac{1}{\mathcal{D}-2}} .
\label{eq:Rpm}
\end{equation}
In horizon quantum mechanics, the ADM mass is promoted to the Hamiltonian, $M\to\hat H$, and two horizons correspond to two operators $\hat r_\pm$, with $\hat r_\pm^{\mathcal{D}-2}=G_\mathcal{D}\beta_\pm\hat H/(\mathcal{D}-2)$. For an energy eigenstate, $\hat H|\psi_E\rangle=E|\psi_E\rangle$, one has $\hat r_\pm|\psi_E\rangle=R_\pm(E)|\psi_E\rangle$ with $R_\pm(E)$ given by Eq.~\eqref{eq:Rpm} with $M\to E$,
\begin{equation}
r_\pm(E)=\ell_\mathcal{D}\left[\frac{\beta_\pm E}{(\mathcal{D}-2)\,\Lambda_{\mathcal{D}}}\right]^{1/(\mathcal{D}-2)} .
\label{eq:rpmE}
\end{equation}

For a source with energy \(E\geq m\), the horizon radius \(r_\pm(E)\) is given by Eq.~\eqref{eq:rpmE}. Since the minimum energy of a source is its rest mass \(m\), the corresponding minimum horizon radius is found by setting \(E=m\) in Eq.~\eqref{eq:rpmE}. Thus,
\begin{equation}
	R_{\min,\pm}\equiv r_\pm(m)
	=\ell_\mathcal{D}\left[\frac{\beta_\pm\,m}{(\mathcal{D}-2)\,\Lambda_{\mathcal{D}}}\right]^{\frac{1}{\mathcal{D}-2}}.
	\label{eq:Rmin}
\end{equation}
Therefore, only horizon radii satisfying $r_\pm\geq R_{\min,\pm}$ are allowed. This constraint is incorporated into the HWF via the Heaviside step function $(\Theta(r_\pm-R_{\min,\pm}))$, as shown in Eq.~\eqref{eq:HWF}. A generic quantum source is not in an energy eigenstate but in a superposition of energies. The horizon radius then fluctuates, and each horizon is described by its own probability distribution. The HWF of each branch is obtained from the momentum-space wave function (Eq.~\eqref{eq:momentum_wavefunction}) with $\Delta_p=\hbar/\ell$. The mass-shell relation
$p^2=E^2-m^2$ gives $\tilde\psi_S\propto e^{-\ell^2(E^2-m^2)/2}$, and the constant factor $e^{\ell^2m^2/2}$ is absorbed in the normalisation. Inverting Eq.~\eqref{eq:rpmE} gives $E=(\mathcal{D}-2)\,\Lambda_{\mathcal{D}}\,\beta_\pm^{-1}(r_\pm/\ell_\mathcal{D})^{\mathcal{D}-2}$, and imposing $r_\pm\ge R_{\min,\pm}$ yields
\begin{equation}
\psi_H^\pm(r_\pm)=\mathcal N_\pm\,\Theta(r_\pm-R_{\min,\pm})\,
\exp\!\left[-\frac{\kappa_\pm}{2}\,r_\pm^{2(\mathcal{D}-2)}\right],
\qquad
\kappa_\pm=\frac{(\mathcal{D}-2)^2\ell^2}{\ell_\mathcal{D}^{2(\mathcal{D}-1)}\beta_\pm^2}.
\label{eq:HWF}
\end{equation}
With the scalar product
$\langle\psi_H|\phi_H\rangle=\Omega_{\mathcal{D}-1}\int_0^\infty\psi_H^*\phi_H\,r^{\mathcal{D}-1}dr$ and
$a_\mathcal{D}=\mathcal{D}/[2(\mathcal{D}-2)]$, the normalisation is
\begin{equation}
|\mathcal N_\pm|^2=\frac{2(\mathcal{D}-2)\,\kappa_\pm^{a_\mathcal{D}}}{\Omega_{\mathcal{D}-1}\,\Gamma(a_\mathcal{D},\ell^2m^2)},
\qquad
\kappa_\pm R_{\min,\pm}^{2(\mathcal{D}-2)}=\ell^2m^2,
\label{eq:Npm}
\end{equation}
For $\ell=\ell_\mathcal{D}\Lambda_{\mathcal{D}}/m$ the
argument of the incomplete gamma function equals one, and $\Gamma(a_\mathcal{D},\kappa_\pm R_{\min,\pm}^{2(\mathcal{D}-2)})$ reduces to $\Gamma(a_\mathcal{D},1)$ in units $\hbar=\ell_\mathcal{D}\Lambda_{\mathcal{D}}=1$. After simplification, we get the HWF as,
\begin{equation}
\psi_H^\pm(r_\pm)=\left[\frac{2(\mathcal{D}-2)\kappa_\pm^{a_\mathcal{D}}}{\Omega_{\mathcal{D}-1}\Gamma(a_\mathcal{D},1)}\right]^{1/2}\Theta\!\left(r_\pm - R_{\min,\pm}\right)\exp\!\left[-\frac{\kappa_\pm}{2}r_\pm^{2(\mathcal{D}-2)}\right].
\label{eq:HWFplus}
\end{equation}
The probability of a black hole horizon ($R_\pm$) being at a gravitational radius $r_\pm$ is given by the horizon probability density. This shows how the probability of detecting a horizon depends on its radius. The horizon probability density associated with each horizon branch is given by,
\begin{equation}
P_H^\pm(r_\pm)=\Omega_{\mathcal{D}-1}r_\pm^{\mathcal{D}-1}|\psi_H^\pm(r_\pm)|^2.
\label{eq:PHplus}
\end{equation}
Consequently, the normalised horizon probability density is, 
\begin{equation}
P_H^\pm(r_\pm) =\frac{2(\mathcal{D}-2)\kappa_\pm^{a_\mathcal{D}}}{\Gamma(a_\mathcal{D},1)}r_\pm^{\mathcal{D}-1}e^{-\kappa_\pm{r_\pm^{2(\mathcal{D}-2)}}}\Theta(r_\pm-R_{\min,\pm})
\end{equation}
\section{Formation Probabilities for the TRN Black Holes}
\subsection{Joint Probability Densities}
\label{subsec:joint}
In the HWF framework, we treat the source position and the horizon radius as
statistically independent, so that the joint probability density that the horizon of the branch $\pm$ is located at $r_\pm$, and that the source is localised within that radius is
\begin{equation}
P_J^\pm(r_\pm)=P_H^\pm(r_\pm)\,P_S(r<r_\pm),
\label{eq:PJ}
\end{equation}

\begin{figure}[]
	\centering
	\includegraphics[width=0.8\textwidth]{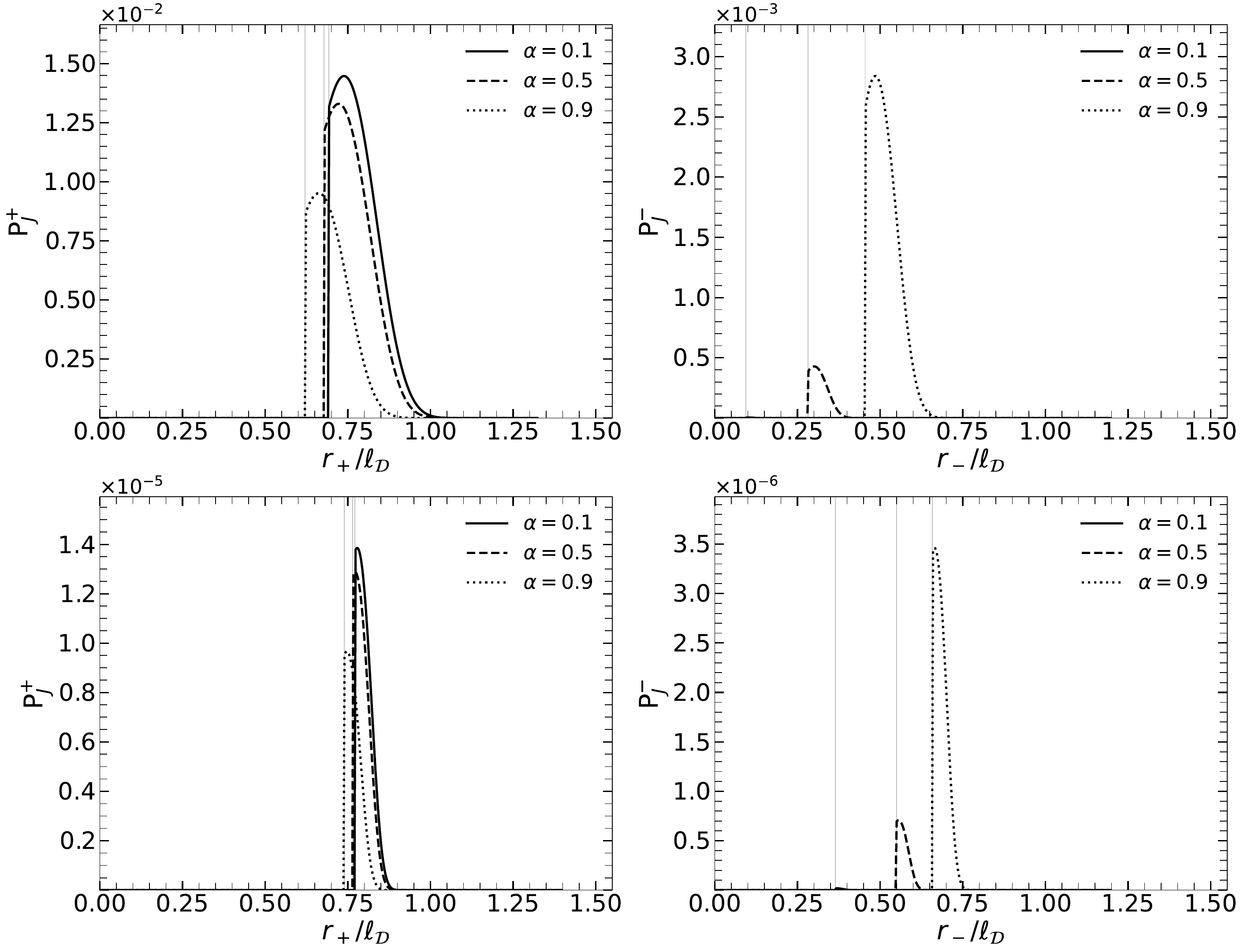}
	\caption{Variation of $P^\pm_{J}$ with horizon radii $r_\pm$ for the $\mathcal{D}=5$ (upper) and $\mathcal{D}=10$ (lower) with $\alpha=0.1$ (solid), $\alpha=0.5$ (dashed), and $\alpha=0.9$ (dotted) at $m/\Lambda_{\mathcal{D}}=0.5$ (or $\ell/\ell_\mathcal{D} = 2$). }
	\label{fig:combined_M0.5_linear} 
\end{figure}

\begin{figure}[]
	\centering
	\includegraphics[width=0.8\textwidth]{ 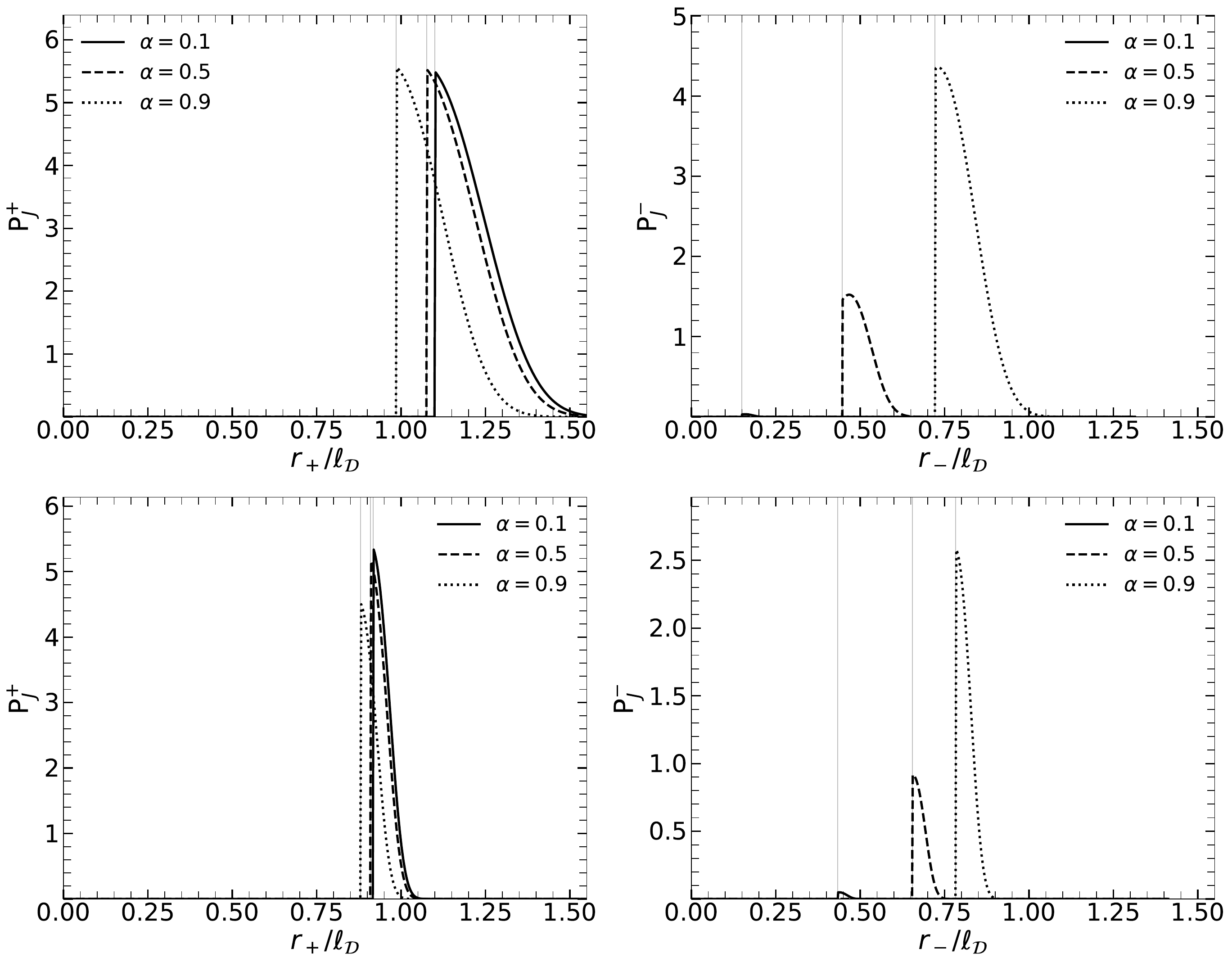}
	\caption{Variation of $P^\pm_{J}$ with horizon radii $r_\pm$ for the $\mathcal{D}=5$ (upper) and $\mathcal{D}=10$ (lower) with $\alpha=0.1$ (solid), $\alpha=0.5$ (dashed), and $\alpha=0.9$ (dotted) at $m/\Lambda_{\mathcal{D}}=2$ (or $\ell/\ell_\mathcal{D} = 0.5$). }
	\label{fig:combined_M2_linear}
\end{figure}

Figures~\ref{fig:combined_M0.5_linear} and~\ref{fig:combined_M2_linear} show $P_J^{\pm}$ for $m=0.5\,\Lambda_{\mathcal{D}}$ and $m=2\,\Lambda_{\mathcal{D}}$ respectively, for $\mathcal{D}= 5$ and $\mathcal{D}=10$. The vertical lines indicate the minimum allowed radius $R_{\min,\pm}$, below which the corresponding densities vanish due to the step function constraint $\Theta(r_\pm-R_{\min,\pm})$. This explains why each curve starts abruptly at $R_{\min,\pm}$.

For $m=0.5\,\Lambda_{\mathcal{D}}$, the $P^+_{J}$ peak shifts towards smaller radii and decreases in height as $\alpha$ increases. The $P^-_{J}$ shows the opposite behaviour, becoming larger at higher values of $\alpha$. These trends are observed for both dimensions. The peaks for $\mathcal{D}=10$ are also much smaller and have a narrower width than those for $\mathcal{D}=5$. The peak height increases strongly with $m$. For $\mathcal{D}=5$ and $\alpha=0.1$, the maximum of $P_J^{+}$ rises from approximately $1.5\times10^{-2}$ at $m=0.5\,\Lambda_{\mathcal{D}}$ (or $\ell =2\ell_\mathcal{D}$) to about $5.4$ at $m=2\,\Lambda_{\mathcal{D}}$ (or $\ell =0.5\ell_\mathcal{D}$), and for $\mathcal{D}=10$ from about $1.4\times10^{-5}$ to about $5.4$. 
These results show that $P_J^\pm(r_\pm)$ decreases when $m$ gets smaller than the $\Lambda_{\mathcal{D}}$ (or equivalently $\ell$ increases above $\ell_\mathcal{D}$). At $m<\Lambda_{\mathcal{D}}$, the peak is much lower in $\mathcal{D}=10$ than in $\mathcal{D}=5$, while at $m=2\,\Lambda_{\mathcal{D}}$ the two peaks are comparable. The $P^-_{J}$ shows the strongest dependence on $\alpha$ than the $P^+_{J}$. At small values of $\alpha$, it is very small, nearly vanishing. As $\alpha$ approaches its extremal value, the $P^-_{J}$ increases and approaches the $P^+_{J}$, indicating the convergence of the two horizons. Similarly, $P^+_{J}$ varies only slightly with $\alpha$, rather than increasing from zero.

\subsection{Numerical Analysis}
\label{subsec:BH_probability}
\begin{figure}[H]
	\centering
	\includegraphics[width=\textwidth]{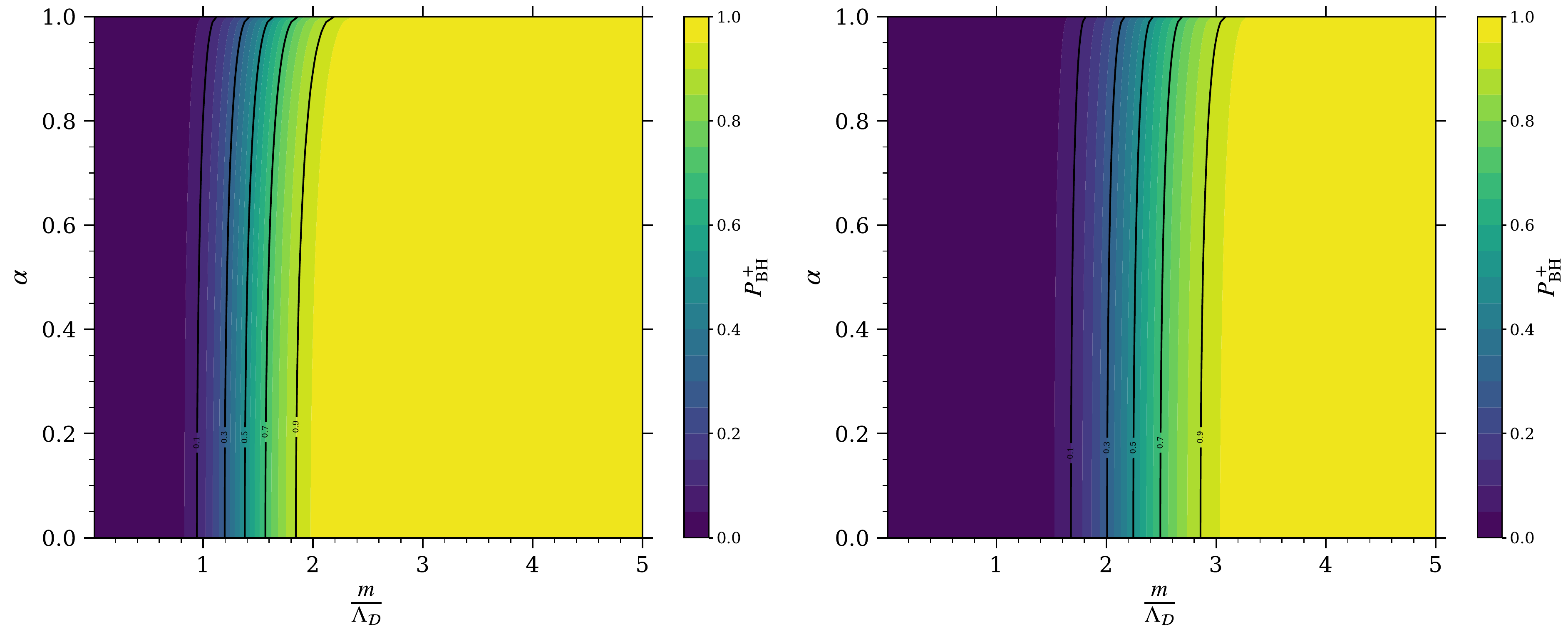}
	\caption{Outer black hole formation probability $P_{\rm BH}^{+}$ in the $(m/\Lambda_{\mathcal{D}},\alpha)$ parameter space for $\mathcal{D}=5$ (left) and $\mathcal{D}=10$ (right).}
	\label{fig:Contour_PBH_outer_mass_alpha}
\end{figure}

\begin{figure}[H]
	\centering
	\includegraphics[width=\textwidth]{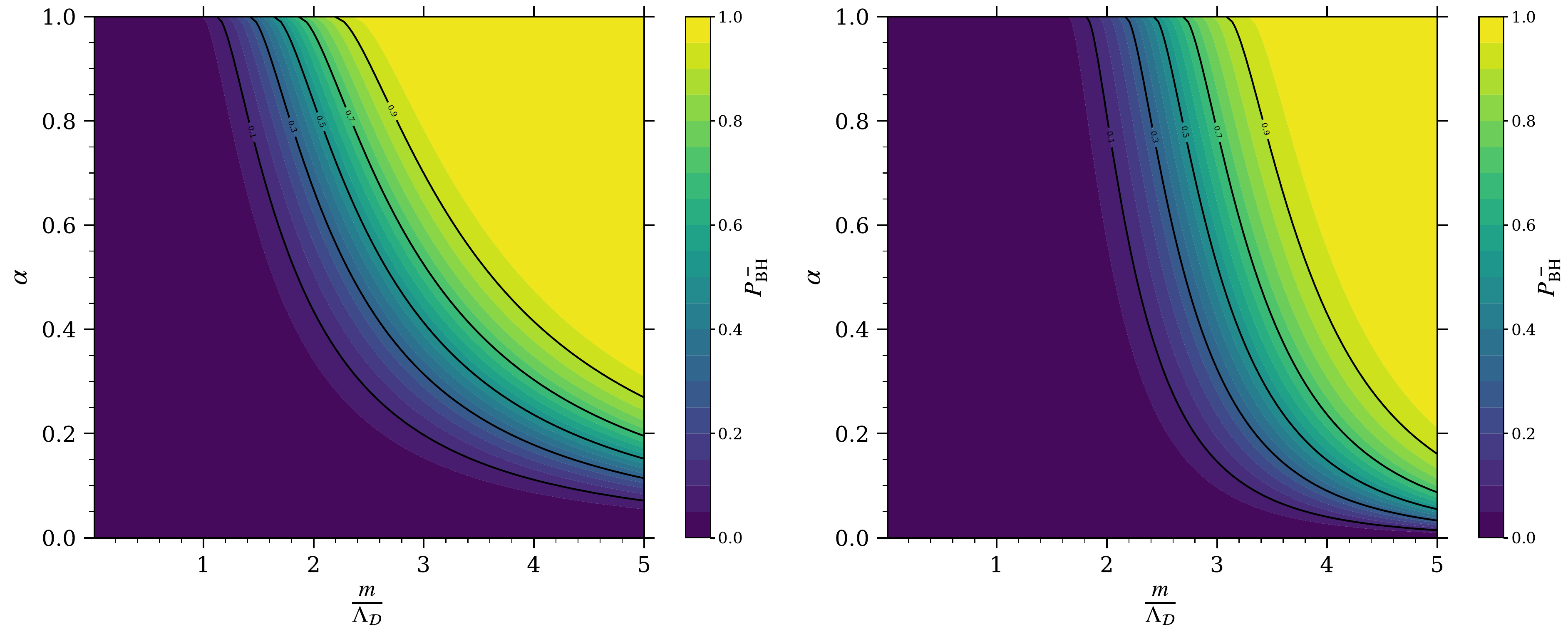}
	\caption{Inner black hole formation probability $P_{\rm BH}^{-}$ in the $(m/\Lambda_{\mathcal{D}},\alpha)$ parameter space for $\mathcal{D}=5$ (left) and $\mathcal{D}=10$ (right).}
	\label{fig:Contour_PBH_inner_mass_alpha} 
\end{figure}

\begin{figure}[H]
	\centering
	\includegraphics[width=\textwidth]{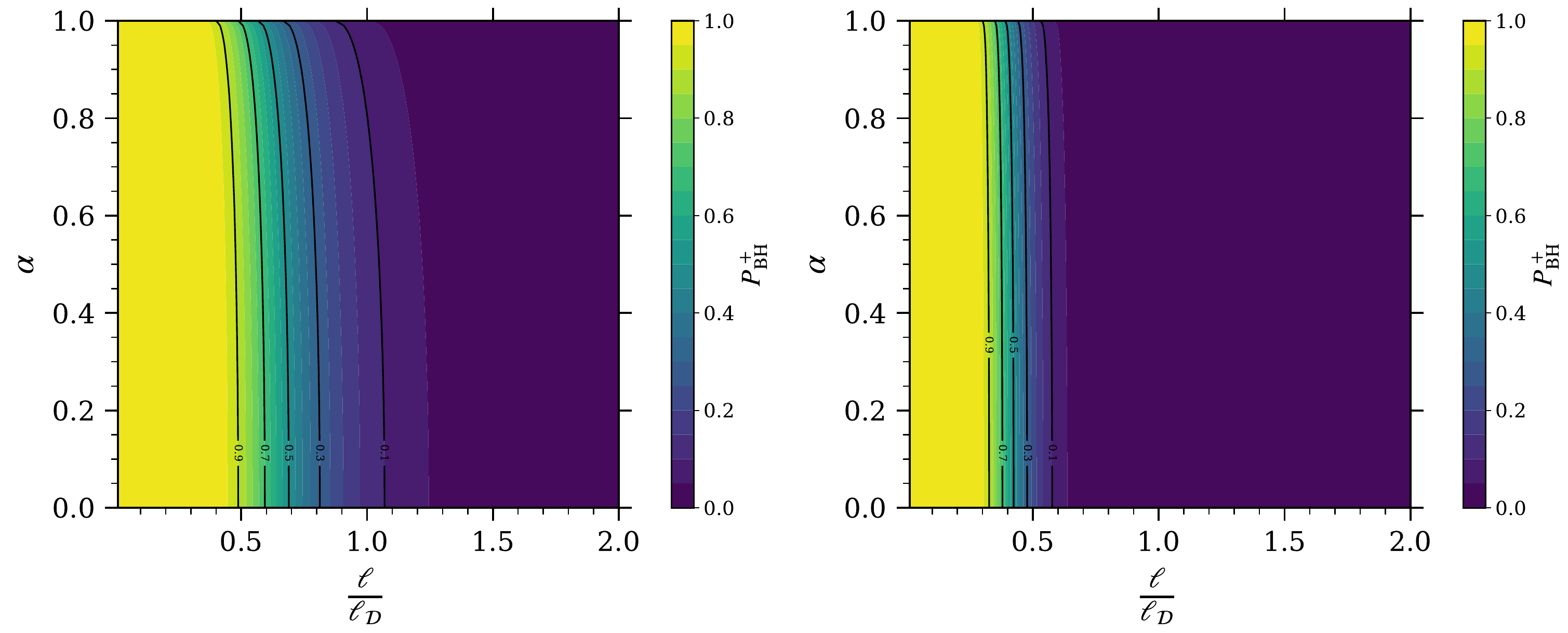}
	\caption{Outer black hole formation probability $P_{\rm BH}^{+}$ in the $(\ell/\ell_\mathcal{D},\alpha)$ parameter space for $\mathcal{D}=5$ (left) and $\mathcal{D}=10$ (right), with $m = \Lambda_{\mathcal{D}}$.}
	\label{fig:Contour_PBH_outer_ell_alpha}
\end{figure}

\begin{figure}[H]
	\centering
	\includegraphics[width=\textwidth]{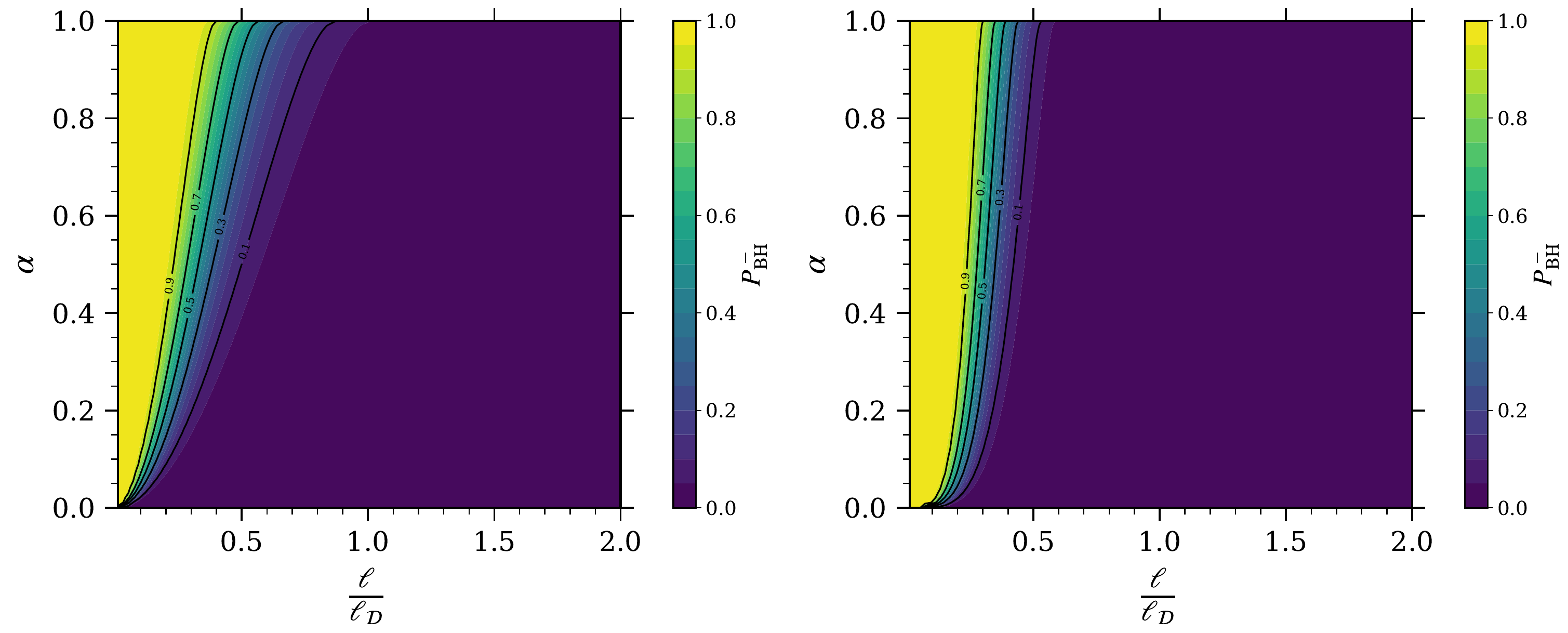}
	\caption{Inner black hole formation probability $P_{\rm BH}^{-}$ in the $(\ell/\ell_\mathcal{D},\alpha)$ parameter space for $\mathcal{D}=5$ (left) and $\mathcal{D}=10$ (right), with $m = \Lambda_{\mathcal{D}}$.}
	\label{fig:Contour_PBH_inner_ell_alpha}
\end{figure}

We assume that a charge source particle, described by a Gaussian wave function, can form a black hole associated with a given horizon branch if it is localised within the corresponding horizon radius $r_\pm$. The probability of black hole formation for each horizon branch is obtained by integrating $P^\pm_{J}$ over the allowed horizon radii,
\begin{equation}
P_{\rm BH}^\pm=\int_{R_{\min,\pm}}^{\infty}P_J^\pm\,dr_\pm ,
\label{eq:PBH-def}
\end{equation}
where the lower limit follows from the step function in Eq.~\eqref{eq:HWF}. The probabilities $P_{\rm BH}^+$ and $P_{\rm BH}^-$ refer to the outer- and inner-horizon branches, respectively. These are distinct, non-exclusive probabilities and must not be interpreted as complementary outcomes. Black hole formation probability for $\mathcal{D}=5$ and $\mathcal{D}=10$ are shown as functions of $m/\Lambda_{\mathcal{D}}$ and $\alpha$ in Figs.~\ref{fig:Contour_PBH_outer_mass_alpha} and~\ref{fig:Contour_PBH_inner_mass_alpha}, and as functions of $\ell/\ell_\mathcal{D}$ and $\alpha$ in Figs.~\ref{fig:Contour_PBH_outer_ell_alpha} and~\ref{fig:Contour_PBH_inner_ell_alpha}. $P_{\rm BH}^+$ increases with $m$ and decreases as $\ell$ increases, so that black hole formation is most likely for heavy, strongly localised sources. Furthermore, the probability $P_{\rm BH}^+$ is suppressed as $\mathcal{D}$ increases. For example, for $\alpha=0.1$ and $m=2\,\Lambda_{\mathcal{D}}$ (or $\ell/\ell_\mathcal{D}=0.5$), $P_{\rm BH}^+=0.956$ is obtained for  $\mathcal{D}=5$, but $0.294$ for $\mathcal{D}=10$ (Table~\ref{tab:PBH}). Similarly, for $\alpha=0.5$, $P_{\rm BH}^+$ exceeds $0.9$ at $m\simeq2\,\Lambda_{\mathcal{D}}$ for $\mathcal{D}=5$, but only at $m\simeq3\,\Lambda_{\mathcal{D}}$ for $\mathcal{D}=10$. Therefore, a larger $m/\Lambda_{\mathcal{D}}$, or equivalently a smaller $\ell/\ell_\mathcal{D}$, is needed in higher dimensions to reach a comparable probability. The two horizon branches behave differently. $P_{\rm BH}^+$ rises earlier with $m$ than $P_{\rm BH}^-$, which lags behind; For example, at $\mathcal{D}=5$, $\alpha=0.5$ and $m=2\,\Lambda_{\mathcal{D}}$, $P_{\rm BH}^+=0.95$ while $P_{\rm BH}^-=0.15$. The effect of $\alpha$ is also different. $P_{\rm BH}^+$ slightly changes with $\alpha$, whereas $P_{\rm BH}^-$ grows by several orders of magnitude between $\alpha=0.1$ and $\alpha=0.5$ (e.g., from $1.2\times10^{-5}$ to $2.6\times10^{-3}$ for $\mathcal{D}=5$ and $m=\Lambda_{\mathcal{D}}$). However, $P_{\rm BH}^-$ increases the most at large $\alpha$, where it approaches $P_{\rm BH}^+$. These are summarised in Table~\ref{tab:PBH}.
To quantify the asymmetry between two branches, we define the difference in their black hole formation probabilities as, 
\begin{equation}
\Delta P_{\rm BH}
=
P_{\rm BH}^{+}-P_{\rm BH}^{-}.
\label{eq:Delta_PBH}
\end{equation}

\begin{figure}[H]
	\centering
	\includegraphics[width=0.80\textwidth]{ 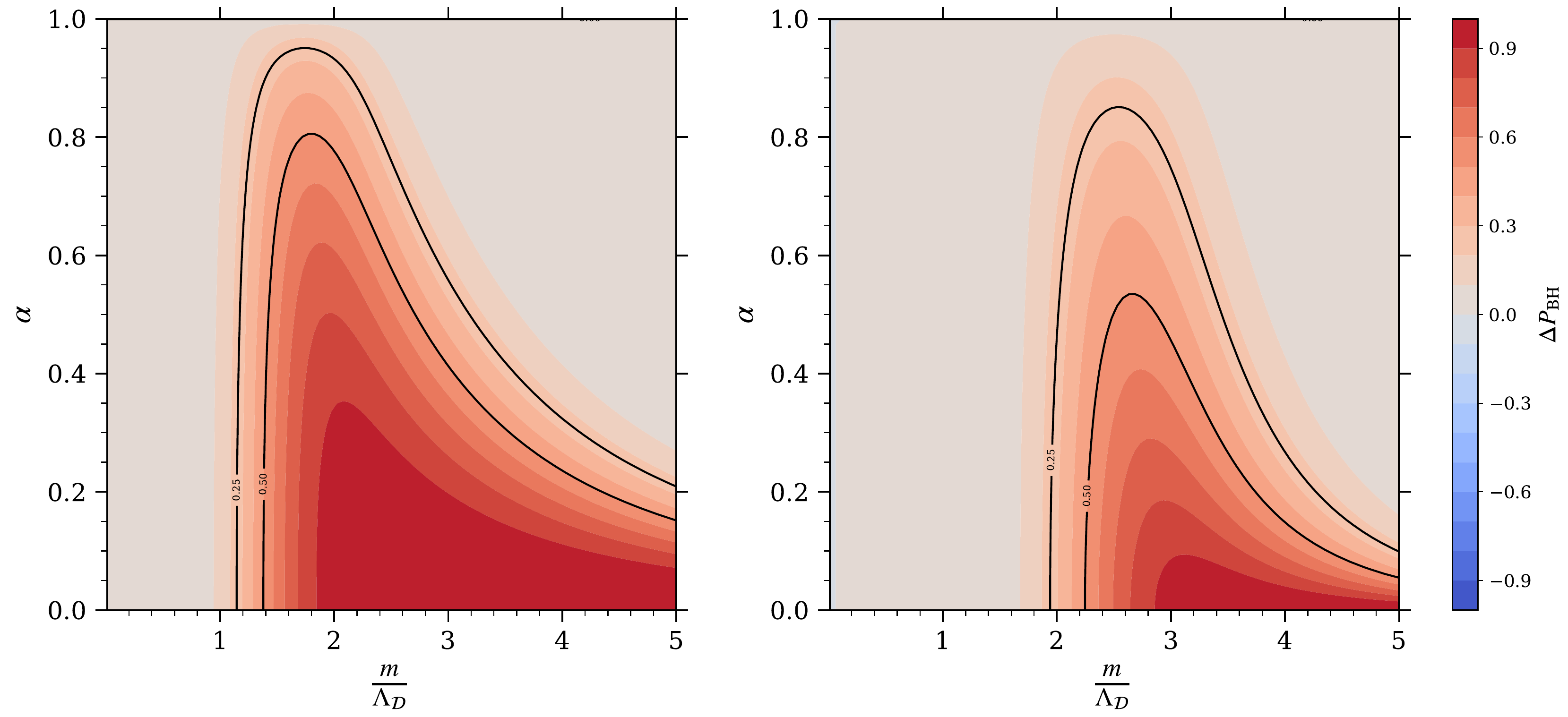}
	\caption{Variation of the asymmetry ($\Delta P_{\rm BH}$) with respect to $(m/\Lambda_{\mathcal{D}}$, $\alpha$) parameter space for $\mathcal{D}=5$ and $\mathcal{D}=10$, respectively. }
	\label{fig:Contour_PBH_difference_mass_alpha}
\end{figure}

Figures~\ref{fig:Contour_PBH_difference_mass_alpha} shows this difference between $P_{\rm BH}^{+}$ and $P_{\rm BH}^{-}$ in the $(m/\Lambda_{\mathcal{D}},\alpha)$ parameter space for $\mathcal{D}=5$ and $\mathcal{D}=10$. Positive values indicate that the outer-horizon probability exceeds the inner-horizon probability ($\Delta P_{\rm BH}>0$), while negative values indicate the opposite ordering. The difference is predominantly positive and is largest at low $\alpha$ and intermediate masses. As $\alpha$ increases, the difference decreases, while increasing $\mathcal{D}$ suppresses its maximum and shifts it towards higher masses. At fixed mass, $\Delta P_{\rm BH}$ decreases with increasing $\alpha$. For $m=2\Lambda_{\mathcal{D}}$, it decreases from $0.954$ to $0.801$ and $0.311$ for $\mathcal{D}=5$, and from $0.292$ to $0.243$ and $0.114$ for $\mathcal{D}=10$, as $\alpha$ increases from $0.1$ to $0.5$ and $0.9$, respectively. At larger
masses, the difference gradually decreases as the two branch probabilities approach one another. In the extremal limit, $\alpha\to1$, the horizon radii coincide and $\Delta P_{\rm BH}\to0$.

\begin{figure}[H]
	\centering
	\includegraphics[width=0.90\textwidth]{ 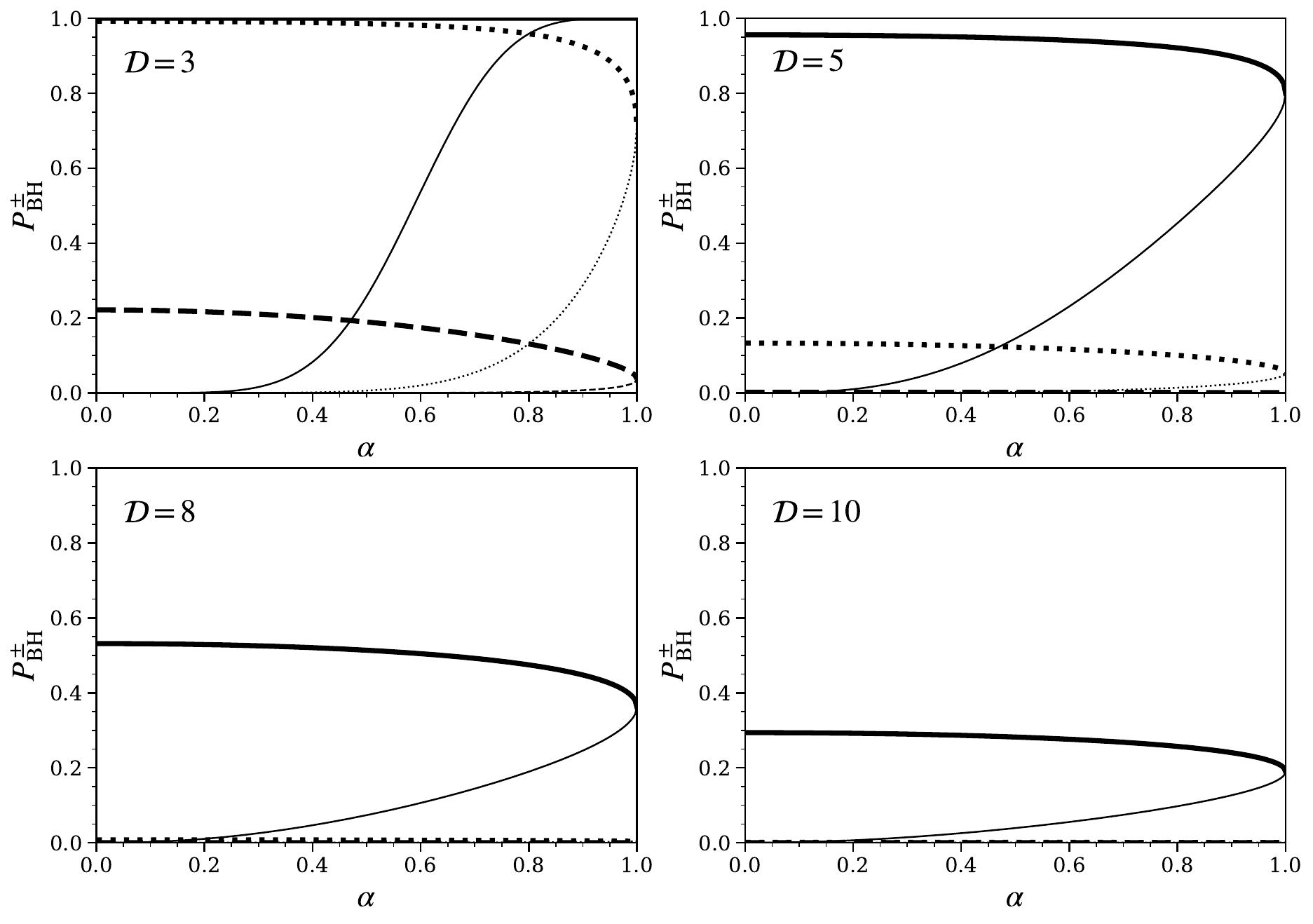}
	\caption{Probabilities $P_{\rm BH}^{\pm}$ as functions of $\alpha$ for $\mathcal{D}=3$, $5$, $8$ and $10$. Thick and thin curves represent $P_{\rm BH}^{+}$ and $P_{\rm BH}^{-}$, respectively, at $m/\Lambda_{\mathcal{D}}=0.5$ (dashed), $1$ (dotted) and $2$ (solid).}
	\label{fig:PBH_both_branches_vs_alpha_final}
\end{figure}
The dependence of both probabilities on $\alpha$ is further illustrated in Figs.~\ref{fig:PBH_both_branches_vs_alpha_final} for $\mathcal{D}=3$, $5$, $8$, and $10$, and for $m/\Lambda_{\mathcal{D}}=0.5$, $1$, and $2$. For $\mathcal{D}=3$, the results in Figs.~\ref{fig:PBH_both_branches_vs_alpha_final} are consistent with those reported in~\cite{Casadio:2015rwa}. $P_{\rm BH}^{+}$ exceeds $P_{\rm BH}^{-}$ at small $\alpha$, whereas the $P_{\rm BH}^{-}$ increases as the charge parameter approaches extremality. The overall suppression observed at higher $\mathcal{D}$ is consistent with the contour plots.

\begin{table}[H]
	\centering
	\caption{Horizon radii ($r_\pm$) and black hole formation probabilities ($P^\pm_{\mathrm{BH}}$) for various values of $\mathcal{D}$, $m/\Lambda_{\mathcal{D}}$ (or equivalently, $\ell/\ell_{\mathcal{D}}$), and $\alpha$.}
	\label{tab:PBH}
	\scriptsize
	\setlength{\tabcolsep}{3pt}
	
	\resizebox{\textwidth}{!}{%
		\begin{tabular}{cccc}
			\toprule
			$m/\Lambda_{\mathcal{D}}$ ($\ell/\ell_{\mathcal{D}}$)
			& $\alpha$
			& $(R_+/\ell_{\mathcal{D}},\,R_-/\ell_{\mathcal{D}})$
			& $\left(
			-\log_{10}P_{\mathrm{BH}}^+,\,
			-\log_{10}P_{\mathrm{BH}}^-,\,
			\log_{10}\frac{P_{\mathrm{BH}}^+}{P_{\mathrm{BH}}^-}
			\right)$ \\
						\toprule
			\multicolumn{4}{c}{$\mathcal{D}=3$} \\
			\midrule
			
			\multirow{3}{*}{0.5 (2.0)}
			& 0.1 & $(0.997,\,0.003)$ & $(0.656,\,8.270,\,7.614)$ \\
			& 0.5 & $(0.933,\,0.067)$ & $(0.722,\,3.990,\,3.268)$ \\
			& 0.9 & $(0.718,\,0.282)$ & $(1,~~~~~\,2.133,\,1.133)$ \\
			
			\cmidrule(lr){1-4}
			
			\multirow{3}{*}{1.0 (1.0)}
			& 0.1 & $(1.995,\,0.005)$ & $(0.004,\,6.464,\,6.461)$ \\
			& 0.5 & $(1.866,\,0.134)$ & $(0.006,\,2.198,\,2.192)$ \\
			& 0.9 & $(1.436,\,0.564)$ & $(0.034,\,0.541,\,0.507)$ \\
			
			\cmidrule(lr){1-4}
			
			\multirow{3}{*}{2.0 (0.5)}
			& 0.1 & $(3.99,~~\,0.01)$ & $(0,~~~~\,4.658,\,4.658)$ \\
			& 0.5 & $(3.732,\,0.268)$ & $(0,~~~~\,0.588,\,0.588)$ \\
			& 0.9 & $(2.872,\,1.128)$ & $(0,~~~~\,0.001,\,0.001)$ \\
			
			\midrule
			\multicolumn{4}{c}{$\mathcal{D}=5$} \\
			\midrule
			
			\multirow{3}{*}{0.5 (2.0)}
			& 0.1 & $(0.693,\,0.094)$ & $(2.638,\,6.925,\,4.286)$ \\
			& 0.5 & $(0.678,\,0.282)$ & $(2.685,\,4.554,\,1.869)$ \\
			& 0.9 & $(0.621,\,0.455)$ & $(2.867,\,3.526,\,0.659)$ \\
			
			\cmidrule(lr){1-4}
			
			\multirow{3}{*}{1.0 (1.0)}
			& 0.1 & $(0.873,\,0.119)$ & $(0.876,\,4.923,\,4.047)$ \\
			& 0.5 & $(0.854,\,0.355)$ & $(0.912,\,2.589,\,1.677)$ \\
			& 0.9 & $(0.782,\,0.573)$ & $(1.059,\,1.627,\,0.569)$ \\
			
			\cmidrule(lr){1-4}
			
			\multirow{3}{*}{2.0 (0.5)}
			& 0.1 & $(1.1,~~~\,0.15)$ & $(0.0198,\,2.946,\,2.926)$ \\
			& 0.5 & $(1.075,\,0.447)$ & $(0.0239,\,0.838,\,0.814)$ \\
			& 0.9 & $(0.986,\,0.722)$ & $(0.0462,\,0.231,\,0.184)$ \\
			
			\midrule
			\multicolumn{4}{c}{$\mathcal{D}=10$} \\
			\midrule
			
			\multirow{3}{*}{0.5 (2.0)}
			& 0.1 & $(0.771,\,0.365)$ & $(6.119,\,9.323,\,3.205)$ \\
			& 0.5 & $(0.764,~\,0.55)$ & $(6.154,\,7.556,\,1.402)$ \\
			& 0.9 & $(0.74,~\,0.658)$ & $(6.293,\,6.789,\,0.496)$ \\
			
			\cmidrule(lr){1-4}
			
			\multirow{3}{*}{1.0 (1.0)}
			& 0.1 & $(0.841,\,0.398)$ & $(2.949,\,5.986,\,3.037)$ \\
			& 0.5 & $(0.834,~~\,0.6)$ & $(2.98~,\,4.281,~~\,1.3)$ \\
			& 0.9 & $(0.807,\,0.718)$ & $(3.106,\,3.561,\,0.455)$ \\
			
			\cmidrule(lr){1-4}
			
			\multirow{3}{*}{2.0 (0.5)}
			& 0.1 & $(0.917,\,0.434)$ & $(0.532,\,2.830,\,2.297)$ \\
			& 0.5 & $(0.909,\,0.654)$ & $(0.55,~\,1.408,\,0.858)$ \\
			& 0.9 & $(0.88,~\,0.783)$ & $(0.619,\,0.898,\,0.279)$ \\
			
			\bottomrule
		\end{tabular}%
	}
\end{table}
\clearpage
\section{Summary}
\label{sec:summary}

We extended the horizon wave function formalism to a charged, spherically symmetric source in $\mathcal{D} + 1$ dimensional Tangherlini-Reisner-Nordström (TRN) geometry. The horizon- wave functions associated with the outer- and inner-horizons, $\psi_H^+$ and $\psi_H^-$ were constructed in the aforementioned framework. These probabilities were combined with the probability of the source under consideration to lie within the respective horizon radii in order to obtain the joint probability densities $P_J^\pm$ respectively for inner as well as outer-horizon. The corresponding black hole formation probabilities $P_{\rm BH}^+$ and $P_{\rm BH}^-$ were obtained by integrating $P_J^\pm$ over $r_{\pm}$. These were ultimately analysed in the context of various parameters namely, $m$ (or equivalently $\ell$), $\alpha$ and $\mathcal{D}$. These were presented as Figs. ~\labelcref{fig:combined_M0.5_linear,fig:combined_M2_linear,fig:Contour_PBH_outer_mass_alpha,fig:Contour_PBH_inner_mass_alpha,fig:Contour_PBH_outer_ell_alpha,fig:Contour_PBH_inner_ell_alpha,fig:Contour_PBH_difference_mass_alpha,fig:PBH_both_branches_vs_alpha_final} and Table~\ref{tab:PBH}. Our results show that joint densities peak sharply just above the minimum radius $R_{\min,\pm}$, so the horizon of each branch is confined to a narrow range of radii, narrower in higher $\mathcal{D}$. For both horizons, $P_{\rm BH}^{\pm}$ increases with $m$ and decreases with $\ell$, so that formation is most likely for heavy, strongly localised sources. The two horizons behave differently. At small $\alpha$, the outer-horizon probability is much higher than the inner-horizon one because the inner radius, which scales with $\beta_-\to0$, is near the origin and is not resolved by sources with finite width. As $\alpha$ increases, $P_{\rm BH}^{-}$ grows and the two probabilities converge as $\alpha\to1$, where the horizons coincide. The difference between the two probabilities, $\Delta P_{\rm BH}$, remains positive over the parameter range considered and approaches zero in the extremal limit. Its magnitude is generally larger at a smaller $\alpha$ and decreases as $\alpha$ approaches extremality. For example, at $m=2\Lambda_{\mathcal{D}}$ and $\alpha=0.5$, we obtain $\Delta P_{\rm BH}\simeq0.801$ for $\mathcal{D}=5$ and $\Delta P_{\rm BH}\simeq0.243$ for $\mathcal{D}=10$. These results show that the distinction between the probabilities of outer- and inner-horizon formation depends strongly on both $\alpha$ and $\mathcal{D}$. Further, the $\mathcal{D}$ dimensions also have a significant impact on the overall formation probability. At fixed $m$ and $\alpha$, the outer-horizon probability decreases significantly as $\mathcal{D}$ increases. For $m=\Lambda_{\mathcal{D}}$ and $\alpha=0.5$, the outer-horizon probability is $P_{\rm BH}^{+}\simeq0.122$ for $\mathcal{D}=5$ to $1.05\times10^{-3}$ for $\mathcal{D}=10$, and for $m=0.5\Lambda_{\mathcal{D}}$ it drops from approximately $2.07\times10^{-3}$ to $7.01\times10^{-7}$. These results indicate that a larger mass relative to the reduced gravity scale is generally required to achieve a substantial probability of black hole formation in higher spatial-dimensions. Clearly, these results will have significant implications upon observations of such black holes in experiments.


\begin{thebibliography}{9}
\bibitem{Bel:2007zz}
L.~Bel,
[arXiv:0709.2257 [gr-qc]].

\bibitem{Schwarzschild:1916uq}
K.~Schwarzschild,
Sitzungsber. Preuss. Akad. Wiss. Berlin (Math. Phys. ) \textbf{1916}, 189-196 (1916)
[arXiv:physics/9905030 [physics]].

\bibitem{Thorne:1972ji}
K.~S.~Thorne,

\bibitem{Ida:2002hg}
D.~Ida and K.~i.~Nakao,
Phys. Rev. D \textbf{66}, 064026 (2002)
doi:10.1103/PhysRevD.66.064026
[arXiv:gr-qc/0204082 [gr-qc]].

\bibitem{Casadio:2013aua}
R.~Casadio and F.~Scardigli,
Eur. Phys. J. C \textbf{74}, no.1, 2685 (2014)
doi:10.1140/epjc/s10052-013-2685-2
[arXiv:1306.5298 [gr-qc]].

\bibitem{Dvali:2010bf}
G.~Dvali and C.~Gomez,
[arXiv:1005.3497 [hep-th]].

\bibitem{Dvali:2011aa}
G.~Dvali and C.~Gomez,
Fortsch. Phys. \textbf{61}, 742-767 (2013)
doi:10.1002/prop.201300001
[arXiv:1112.3359 [hep-th]].


\bibitem{Adler:1999bu}
R.~J.~Adler and D.~I.~Santiago,
Mod. Phys. Lett. A \textbf{14}, 1371 (1999)
doi:10.1142/S0217732399001462
[arXiv:gr-qc/9904026 [gr-qc]].

\bibitem{Scardigli:1999jh}
F.~Scardigli,
Phys. Lett. B \textbf{452}, 39-44 (1999)
doi:10.1016/S0370-2693(99)00167-7
[arXiv:hep-th/9904025 [hep-th]].

\bibitem{Hossenfelder:2012jw}
S.~Hossenfelder,
Living Rev. Rel. \textbf{16}, 2 (2013)
doi:10.12942/lrr-2013-2
[arXiv:1203.6191 [gr-qc]].

\bibitem{Alberghi:2013hca}
G.~L.~Alberghi, L.~Bellagamba, X.~Calmet, R.~Casadio and O.~Micu,
Eur. Phys. J. C \textbf{73}, no.6, 2448 (2013)
doi:10.1140/epjc/s10052-013-2448-0
[arXiv:1303.3150 [hep-ph]].


\bibitem{Giddings:2004xy}
S.~B.~Giddings and V.~S.~Rychkov,
Phys. Rev. D \textbf{70}, 104026 (2004)
doi:10.1103/PhysRevD.70.104026
[arXiv:hep-th/0409131 [hep-th]].

\bibitem{Casadio:2013tma}
R.~Casadio,
Eur. Phys. J. Plus \textbf{139}, no.8, 770 (2024)
doi:10.1140/epjp/s13360-024-05575-4
[arXiv:1305.3195 [gr-qc]].

\bibitem{Casadio:2013uga}
R.~Casadio, O.~Micu and F.~Scardigli,
Phys. Lett. B \textbf{732}, 105-109 (2014)
doi:10.1016/j.physletb.2014.03.037
[arXiv:1311.5698 [hep-th]].


\bibitem{Tangherlini:1963bw}
F.~R.~Tangherlini,
Nuovo Cim. \textbf{27}, 636-651 (1963)
doi:10.1007/BF02784569

\bibitem{Myers:1986un}
R.~C.~Myers and M.~J.~Perry,
Annals Phys. \textbf{172}, 304 (1986)
doi:10.1016/0003-4916(86)90186-7

\bibitem{Emparan:2008eg}
R.~Emparan and H.~S.~Reall,
Living Rev. Rel. \textbf{11}, 6 (2008)
doi:10.12942/lrr-2008-6
[arXiv:0801.3471 [hep-th]].

\bibitem{Tomizawa:2011mc}
S.~Tomizawa and H.~Ishihara,
Prog. Theor. Phys. Suppl. \textbf{189}, 7-51 (2011)
doi:10.1143/PTPS.189.7
[arXiv:1104.1468 [hep-th]].

\bibitem{Meschini:2006gm}
D.~Meschini,
Found. Sci. \textbf{12}, 277-294 (2007)
doi:10.1007/s10699-006-9102-3
[arXiv:gr-qc/0601097 [gr-qc]].


\bibitem{Antoniadis:1998ig}
I.~Antoniadis, N.~Arkani-Hamed, S.~Dimopoulos and G.~R.~Dvali,
Phys. Lett. B \textbf{436}, 257-263 (1998)
doi:10.1016/S0370-2693(98)00860-0
[arXiv:hep-ph/9804398 [hep-ph]].
\bibitem{Arkani-Hamed:1998jmv}
N.~Arkani-Hamed, S.~Dimopoulos and G.~R.~Dvali,
Phys. Lett. B \textbf{429}, 263-272 (1998)
doi:10.1016/S0370-2693(98)00466-3
[arXiv:hep-ph/9803315 [hep-ph]].


\bibitem{Randall:1999ee}
L.~Randall and R.~Sundrum,
Phys. Rev. Lett. \textbf{83}, 3370-3373 (1999)
doi:10.1103/PhysRevLett.83.3370
[arXiv:hep-ph/9905221 [hep-ph]].
\bibitem{Randall:1999vf}
L.~Randall and R.~Sundrum,
Phys. Rev. Lett. \textbf{83}, 4690-4693 (1999)
doi:10.1103/PhysRevLett.83.4690
[arXiv:hep-th/9906064 [hep-th]].


\bibitem{Argyres:1998qn}
P.~C.~Argyres, S.~Dimopoulos and J.~March-Russell,
Phys. Lett. B \textbf{441}, 96-104 (1998)
doi:10.1016/S0370-2693(98)01184-8
[arXiv:hep-th/9808138 [hep-th]].

\bibitem{Banks:1999gd}
T.~Banks and W.~Fischler,
[arXiv:hep-th/9906038 [hep-th]].

\bibitem{Dimopoulos:2001hw}
S.~Dimopoulos and G.~L.~Landsberg,
Phys. Rev. Lett. \textbf{87}, 161602 (2001)
doi:10.1103/PhysRevLett.87.161602
[arXiv:hep-ph/0106295 [hep-ph]].

 \bibitem{Gingrich:2007fm}
D.~M.~Gingrich,
Int. J. Mod. Phys. A \textbf{22}, 5685-5699 (2007)
doi:10.1142/S0217751X0703892X
[arXiv:0706.0627 [hep-ph]].

\bibitem{Giddings:2001ih}
S.~B.~Giddings,
eConf \textbf{C010630}, P328 (2001)
[arXiv:hep-ph/0110127 [hep-ph]].

\bibitem{Kanti:2004nr}
P.~Kanti,
Int. J. Mod. Phys. A \textbf{19}, 4899-4951 (2004)
doi:10.1142/S0217751X04018324
[arXiv:hep-ph/0402168 [hep-ph]].

\bibitem{Cavaglia:2002si}
M.~Cavaglia,
Int. J. Mod. Phys. A \textbf{18}, 1843-1882 (2003)
doi:10.1142/S0217751X03013569
[arXiv:hep-ph/0210296 [hep-ph]].

\bibitem{Dimopoulos:2001qe}
S.~Dimopoulos and R.~Emparan,
Phys. Lett. B \textbf{526}, 393-398 (2002)
doi:10.1016/S0370-2693(01)01525-8
[arXiv:hep-ph/0108060 [hep-ph]].

\bibitem{Koch:2005ks}
B.~Koch, M.~Bleicher and S.~Hossenfelder,
JHEP \textbf{10}, 053 (2005)
doi:10.1088/1126-6708/2005/10/053
[arXiv:hep-ph/0507138 [hep-ph]].


\bibitem{Feng:2001ib}
J.~L.~Feng and A.~D.~Shapere,
Phys. Rev. Lett. \textbf{88}, 021303 (2002)
doi:10.1103/PhysRevLett.88.021303
[arXiv:hep-ph/0109106 [hep-ph]].

\bibitem{Anchordoqui:2001cg}
L.~A.~Anchordoqui, J.~L.~Feng, H.~Goldberg and A.~D.~Shapere,
Phys. Rev. D \textbf{65}, 124027 (2002)
doi:10.1103/PhysRevD.65.124027
[arXiv:hep-ph/0112247 [hep-ph]].


\bibitem{Eardley:2002re}
D.~M.~Eardley and S.~B.~Giddings,
Phys. Rev. D \textbf{66}, 044011 (2002)
doi:10.1103/PhysRevD.66.044011
[arXiv:gr-qc/0201034 [gr-qc]].

\bibitem{Yoshino:2002tx}
H.~Yoshino and Y.~Nambu,
Phys. Rev. D \textbf{67}, 024009 (2003)
doi:10.1103/PhysRevD.67.024009
[arXiv:gr-qc/0209003 [gr-qc]].

\bibitem{Yoshino:2005hi}
H.~Yoshino and V.~S.~Rychkov,
Phys. Rev. D \textbf{71}, 104028 (2005)
[erratum: Phys. Rev. D \textbf{77}, 089905 (2008)]
doi:10.1103/PhysRevD.71.104028
[arXiv:hep-th/0503171 [hep-th]].


\bibitem{Voloshin:2001vs}
M.~B.~Voloshin,
Phys. Lett. B \textbf{518}, 137-142 (2001)
doi:10.1016/S0370-2693(01)01061-9
[arXiv:hep-ph/0107119 [hep-ph]].


\bibitem{CMS:2026cgd}
A.~Hayrapetyan \textit{et al.} [CMS],
[arXiv:2604.10732 [hep-ex]].
	
\bibitem{ATLAS:2023vat}
G.~Aad \textit{et al.} [ATLAS],
Phys. Rev. D \textbf{109}, no.3, 032010 (2024)
doi:10.1103/PhysRevD.109.032010
[arXiv:2307.14967 [hep-ex]].
	
\bibitem{ATLAS:2026hzb}
G.~Aad \textit{et al.} [ATLAS],
[arXiv:2604.19495 [hep-ex]].

\bibitem{Ahmad:2026hcy}
A.~Ahmad, S.~K.~Gupta and A.~Ali,
Nucl. Phys. B \textbf{1032}, 117683 (2026)
doi:10.1016/j.nuclphysb.2026.117683
[arXiv:2606.19839 [hep-ph]].


\bibitem{Casadio:2015qaq}
R.~Casadio, A.~Giugno and O.~Micu,
Int. J. Mod. Phys. D \textbf{25}, no.02, 1630006 (2016)
doi:10.1142/S0218271816300068
[arXiv:1512.04071 [hep-th]].

\bibitem{Casadio:2016fev}
R.~Casadio, A.~Giugno and A.~Giusti,
Gen. Rel. Grav. \textbf{49}, no.2, 32 (2017)
doi:10.1007/s10714-017-2198-7
[arXiv:1605.06617 [gr-qc]].

\bibitem{Casadio:2017nfg}
R.~Casadio, A.~Giugno, A.~Giusti and O.~Micu,
Eur. Phys. J. C \textbf{77}, no.5, 322 (2017)
doi:10.1140/epjc/s10052-017-4882-x
[arXiv:1701.05778 [gr-qc]].

\bibitem{Casadio:2018rlc}
R.~Casadio and O.~Micu,
Eur. Phys. J. C \textbf{78}, no.10, 852 (2018)
doi:10.1140/epjc/s10052-018-6326-7
[arXiv:1806.05944 [gr-qc]].

\bibitem{Arsene:2016kvf}
N.~Arsene, R.~Casadio and O.~Micu,
Eur. Phys. J. C \textbf{76}, no.7, 384 (2016)
doi:10.1140/epjc/s10052-016-4228-0
[arXiv:1606.07323 [hep-ph]].

\bibitem{Feng:2025nai}
W.~Feng, A.~Giusti and R.~Casadio,
Eur. Phys. J. Plus \textbf{140}, no.2, 145 (2025)
doi:10.1140/epjp/s13360-025-06065-x
[arXiv:2408.17091 [gr-qc]].

\bibitem{Spallucci:2021ljr}
E.~Spallucci and A.~Smailagic,
Phys. Lett. B \textbf{816}, 136180 (2021)
doi:10.1016/j.physletb.2021.136180
[arXiv:2103.03947 [hep-th]].


\bibitem{Casadio:2015rwa}
R.~Casadio, O.~Micu and D.~Stojkovic,
JHEP \textbf{05}, 096 (2015)
doi:10.1007/JHEP05(2015)096
[arXiv:1503.01888 [gr-qc]].



\bibitem{Casadio:2015jha}
R.~Casadio, R.~T.~Cavalcanti, A.~Giugno and J.~Mureika,
Phys. Lett. B \textbf{760}, 36-44 (2016)
doi:10.1016/j.physletb.2016.06.042
[arXiv:1509.09317 [gr-qc]].



\bibitem{Casadio:2014twa}
R.~Casadio,
Eur. Phys. J. C \textbf{75}, no.4, 160 (2015)
doi:10.1140/epjc/s10052-015-3404-y
[arXiv:1411.5848 [gr-qc]].

\bibitem{Casadio:2013tza}
R.~Casadio,
Springer Proc. Phys. \textbf{170}, 225-231 (2016)
doi:10.1007/978-3-319-20046-0{\_}26
[arXiv:1310.5452 [gr-qc]].

\bibitem{Casadio:2015sda}
R.~Casadio, O.~Micu and D.~Stojkovic,
Phys. Lett. B \textbf{747}, 68-72 (2015)
doi:10.1016/j.physletb.2015.05.053
[arXiv:1503.02858 [gr-qc]].


\end{thebibliography}
\end{document}